\documentclass[aps,prl,twocolumn,showpacs,superscriptaddress,groupedaddress]{revtex4}  % for review and submission
\usepackage{graphicx}  % needed for figures
\usepackage{dcolumn}   % needed for some tables
\usepackage{bm}        % for math
\usepackage{amssymb}   % for math
\usepackage[dvipsnames,usenames]{color}
\usepackage{color}
\usepackage{pdfpages}

\begin{document}

% The following information is for internal review, please remove them for submission
%\widetext
%\leftline{Version xx as of \today}
%\leftline{Primary authors: Joe E. Physics}
%\leftline{To be submitted to (PRL, PRD-RC, PRD, PLB; choose one.)}
%\leftline{Comment to {\tt d0-run2eb-nnn@fnal.gov} by xxx, yyy}
%\centerline{\em D\O\ INTERNAL DOCUMENT -- NOT FOR PUBLIC DISTRIBUTION}

% the following line is for submission, including submission to the arXiv!!
%\hspace{5.2in} \mbox{Fermilab-Pub-04/xxx-E}
\newcommand{\etal}{{\sl et al.}}
\newcommand{\ie}{{\sl i.e.}}
\newcommand{\sto}{SrTiO$_3$} 
\newcommand{\lto}{LaTiO$_3$}
\newcommand{\lao}{LaAlO$_3$}
\newcommand{\tith}{Ti$^{3+}$}
\newcommand{\tif}{Ti$^{4+}$}
\newcommand{\otw}{O$^{2-}$}
\newcommand{\alo}{AlO$ _2 $}
\newcommand{\tio}{TiO$ _2 $}
\newcommand{\eg}{$e_{g}$}
\newcommand{\tg}{$t_{2g}$}
\newcommand{\dxy}{$d_{xy}$}
\newcommand{\dxz}{$d_{xz}$}
\newcommand{\dyz}{$d_{yz}$}
\newcommand{\egp}{$e'_{g}$}
\newcommand{\ag}{$a_{1g}$}
\newcommand{\muB}{$\mu_{\rm B}$}
\newcommand{\ef}{$E_{\rm F}$}
\newcommand{\alao}{$a_{\rm LAO}$}
\newcommand{\asto}{$a_{\rm STO}$}
\newcommand{\nst}{$N_{\rm STO}$}
\newcommand{\lamstn}{(LAO)$_M$/(STO)$_N$}

\title{Massive symmetry breaking in LaAlO$_3$/SrTiO$_3$(111) quantum wells:\\
      a three-orbital, strongly correlated generalization of graphene}
%\input author_list.tex       % D0 authors (remove the first 3 lines
                             % of this file prior to submission, they
                             % contain a time stamp for the authorlist)
                             % (includes institutions and visitors)
\author{David Doennig}
\affiliation{Department of Earth and Environmental Sciences, Section Crystallography and Center of Nanoscience,
University of Munich, Theresienstr. 41, 80333 Munich, Germany}
\author{Warren E. Pickett}
\affiliation{Department of Physics, University of California Davis, One Shields Avenue, Davis, CA 95616, U.S.A.}
\author{Rossitza Pentcheva}
\email{rossitzap@lmu.de}
\affiliation{Department of Earth and Environmental Sciences, Section Crystallography and Center of Nanoscience,
University of Munich, Theresienstr. 41, 80333 Munich, Germany}
\date{\today}

\begin{abstract}
Density functional theory calculations with an on-site Coulomb repulsion term (GGA+$U$ method) reveal competing ground states in (111) oriented (LaAlO$_3$)$_M$/(SrTiO$_3$)$_N$ superlattices with $n$-type interfaces, ranging from spin, orbital polarized, Dirac point Fermi surface to
 charge ordered flat band phases. These are steered by the interplay of $(i)$ Hubbard $U$, $(ii)$ SrTiO$_3$ quantum well thickness and $(iii)$  crystal field
spitting tied to in-plane strain. In the honeycomb lattice bilayer case $N$=2 under tensile strain inversion symmetry breaking drives the system from a ferromagnetic Dirac point (massless Weyl semimetal) to a charge ordered multiferroic (ferromagnetic and
ferroelectric) flat band massive (insulating) phase. With increasing SrTiO$_3$ quantum well thickness an  insulator-to-metal transition occurs.
\end{abstract}

\maketitle

%\section{\label{sec:level1}First-level heading}
% sections are not used for PRL papers

\begin{figure}[b!]
\includegraphics[scale=0.36,angle=270]{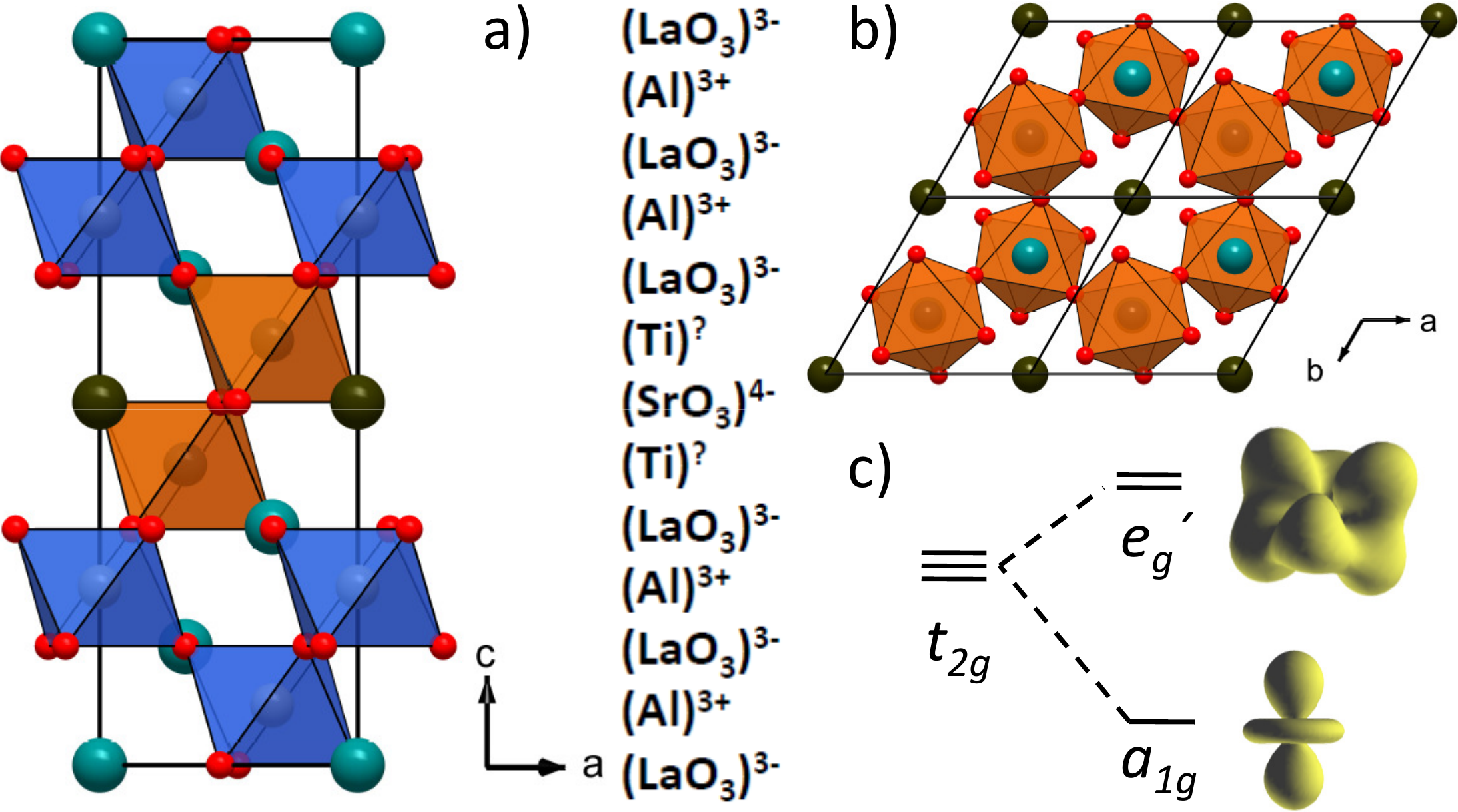}
\caption{\label{fig:struct} a) Side view of (LAO)$_4$/(STO)$_2$(111) SL with an $n$-type interface. b) Top view of the STO bilayer forming a buckled honeycomb lattice out of the two triangular lattices of Ti cations at each interface where Ti are second nearest neighbors. c) splitting of the \tg\ orbitals in \ag\ and \egp\ due to trigonal symmetry, for one sign of the strain.  }
\end{figure}

Remarkably rich electronic behavior has been discovered at oxide interfaces ranging from 
two-dimensional conductivity, superconductivity and magnetism to both confinement induced 
and gate controlled metal-to-insulator 
transitions.\cite{Hwang2012} Most of the interest so far has been directed at (001) 
oriented interfaces as, e.g., the 
ones between the two band insulators \lao\ (LAO) and \sto\ (STO).\cite{Huijben2009,Mannhart,jpcm2010,zubko,overview} Recently 
the growth and initial characterization, including finding of a high mobility electron gas,
of LAO films on STO(111) has been 
reported.\cite{Herranz2012} In contrast 
to the (001) direction where in the perovskite structure  AO and BO$_2$ layers alternate, the (111) 
orientation comprises alternating stacking of AO$_3$ and B layers which can be highly charged: for
example (LaO$_3$)$^{3-}$/Al$^{3+}$ for LAO, (SrO$_3$)$^{4-}$/Ti$^{4+}$ for STO, as illustrated in 
Fig.~\ref{fig:struct}a. Despite the difference in stacking and charge of the individual layers, a polar 
discontinuity arises for both orientations, with a mismatch of $e/2$ per B cation for the 
$n$-type interfaces. For the (001) orientation this polar discontinuity is considered to be the origin 
of the rich spectrum of functional properties mentioned above, albeit the latter can also be influenced 
by defects. It is timely to investigate whether similar electronic reconstructions and exotic phases 
arise for the (111) orientation. 

Perovskite (111) layers have distinctive real space topology: each BO$_6$ layer constitutes 
a triangular lattice where the B cations are {\it second} neighbors. Combining two such layers in 
a bilayer forms a buckled honeycomb lattice, topologically equivalent to that of graphene (Fig.~\ref{fig:struct}b); 
three layers form the also distinctive dice lattice. The possibility for nontrivial topology of electrons 
hopping on a honeycomb lattice proposed by Haldane\cite{Haldane88} has spurred model  Hamiltonians 
studies of topologically nontrivial states for (111)-oriented perovskite 
superlattices,\cite{Xiao2011,Ruegg2011,Yang2011} where the focus was on the LaNiO$_3$ (LNO) $e_g$ system 
confined within LAO with quadratic band touching points, and a Dirac point at higher band 
filling.\cite{Yang2011,Ruegg2011,Ruegg2012} This two-orbital honeycomb lattice is 
beginning to be grown and 
characterized.\cite{Gilbert2012,Middey2012} 

\begin{figure*}[t!]
%\vspace{-5.0cm}
\includegraphics[angle=270,scale=0.7]{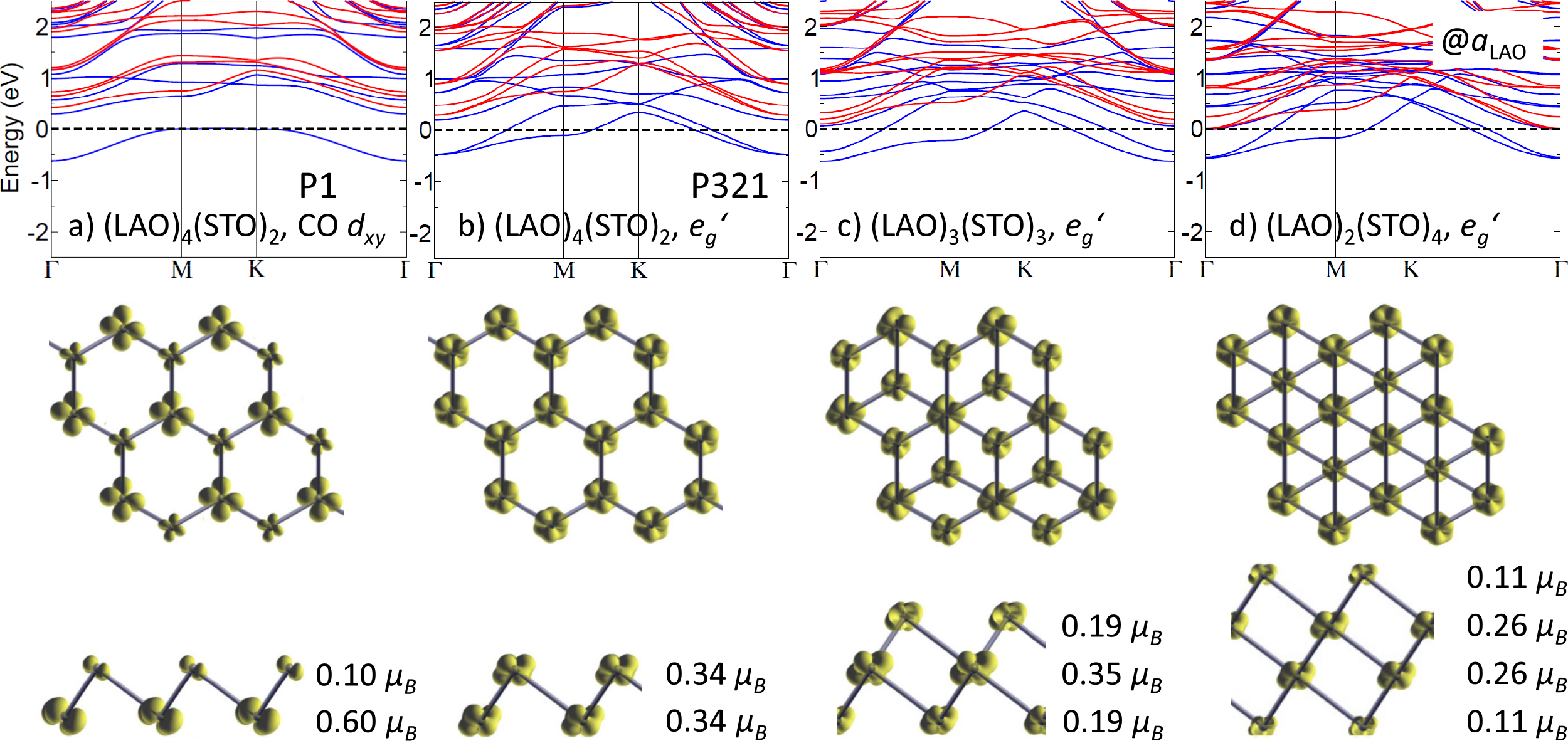}% Here is how to import EPS art
%\vspace{-2.0cm}
\caption{\label{fig:Nsto-alao}Band structure and electron density distribution (top and side view), integrated over occupied Ti $3d$ 
bands, for  \lamstn(111) for the superlattices at \alao. 
Majority and minority bands are plotted in blue and 
red, respectively. a) $N=2$ charge ordered FM insulator.  Note the $d_{xy}$ occupation, {\it i.e.} single 
real $t_{2g}$ orbital. 
b-d) $N=2,3,4$ STO layers, retaining inversion symmetry with \egp\ orbital occupation. For the thicker SLs ($N=3,4$) the excess charge is redistributed from the interface Ti layer to the central
layers, a quantum confinement effect, and the band dispersion at $\Gamma$ increases somewhat. 
}
\end{figure*}

The corresponding three-orbital $t_{2g}$ system is realized for STO
confined in LAO, where $e/2$ charge from each $n$-type ``interface'' (IF) will force one electron 
into 12 Ti conduction states (2 atoms, 3 orbitals, 2 spins), initially with P321 symmetry with two 
generators (3-fold rotation, $z$ mirror+$x$$\leftrightarrow$$y$, which we heuristically refer to as inversion).  
The result is a 12-band, potentially strongly correlated generalization of graphene subject to numerous
symmetry breaking orders: charge, spin, orbital, rotation, inversion ${\cal I}$, time reversal ${\cal T}$,  and
gauge symmetry; our methods do not address the latter. The corresponding model Hamiltonian would include the
symmetry group SU(2)$_{spin}$$\times$SU(2)$_{orb}$$\times$P3$\times$${\cal I}$$\times$${\cal T}$$\times$U(1)$_{gauge}$.

%%\begin{figure}[h!]
%\vspace{-3.0cm}
%%\includegraphics[scale=0.70]{bild_04.pdf}
%\vspace{-3.0cm}
%%\caption{\label{fig:struct} a) Bands and side view of 2STO/4LAO(001) [not (111)] SL with tensile strain. b) Bands and
%%top view, for compressive strain. Although both are FM semimetals with slightly 
%%overlapping but disjoint valence and conduction band complexes, the orbital orderings are {\it opposite}.}
%%\end{figure} 

A key question is that of orbital polarization, which is a primary factor in magnetic, transport, and
optical properties. The geometry of the 111-superlattice breaks orbital 3-fold (\tg) symmetry into trigonal \tg\ $\rightarrow$ \egp\ + \ag\,
as shown schematically in Fig.~\ref{fig:struct}c. 
%(to an undetermined extent) 
For the (001) IF, previous DFT studies
predicted,\cite{pentcheva2006,pentcheva2008,zong2008,popovich} and XAS data\cite{XASdata}
demonstrated, that the \tg\ degeneracy is lifted such that the \dxy\ orbital at the interface 
lies lower in energy. Including static local correlation effects within GGA+$U$ stabilizes a charge ordered
and orbitally polarized layer with alternating \tith\ and \tif\ in the interface layer and a \dxy\ orbital occupied 
at the \tith\ sites.\cite{pentcheva2006,pentcheva2008} It will be instructive to compare this scenario with the
behavior for (111) orientation.
%This scenario is still the only viable
%one to account for the observed insulating ground state, and  

A mathematically symmetric expression adapted to trigonal symmetry for \tg\ orbitals is
\begin{center} 
 $|\psi_m\rangle = (\zeta^0_m |d_{xy}\rangle+\zeta^1_m |d_{yz}\rangle+\zeta_m^2 |d_{xz}\rangle)/\sqrt{3}$,\\
\end{center}
where $\zeta_m = e^{2\pi i m/3}$. One issue is whether complex \egp\ orbitals $m$=1,2 ($m$=0 is the
\ag\ orbital) assert themselves,
inviting anomalously large response to spin-orbit coupling in $t_{2g}$ systems,\cite{SVOSTO_Pardo} 
or whether real combinations of the \egp\ orbitals persist. 
Complex orbitals in the \eg\ bilayer have been predicted to encourage topological phases.\cite{Ruegg2012}
In this paper we find that trigonal level splitting, which is directly connected to strain, 
determines the orbital occupation that vastly influences the
electronic structure in (111) oriented STO quantum well (QW).

\begin{figure*}[tb!]
%\vspace{-5.0cm}
\includegraphics[angle=270,scale=0.7]{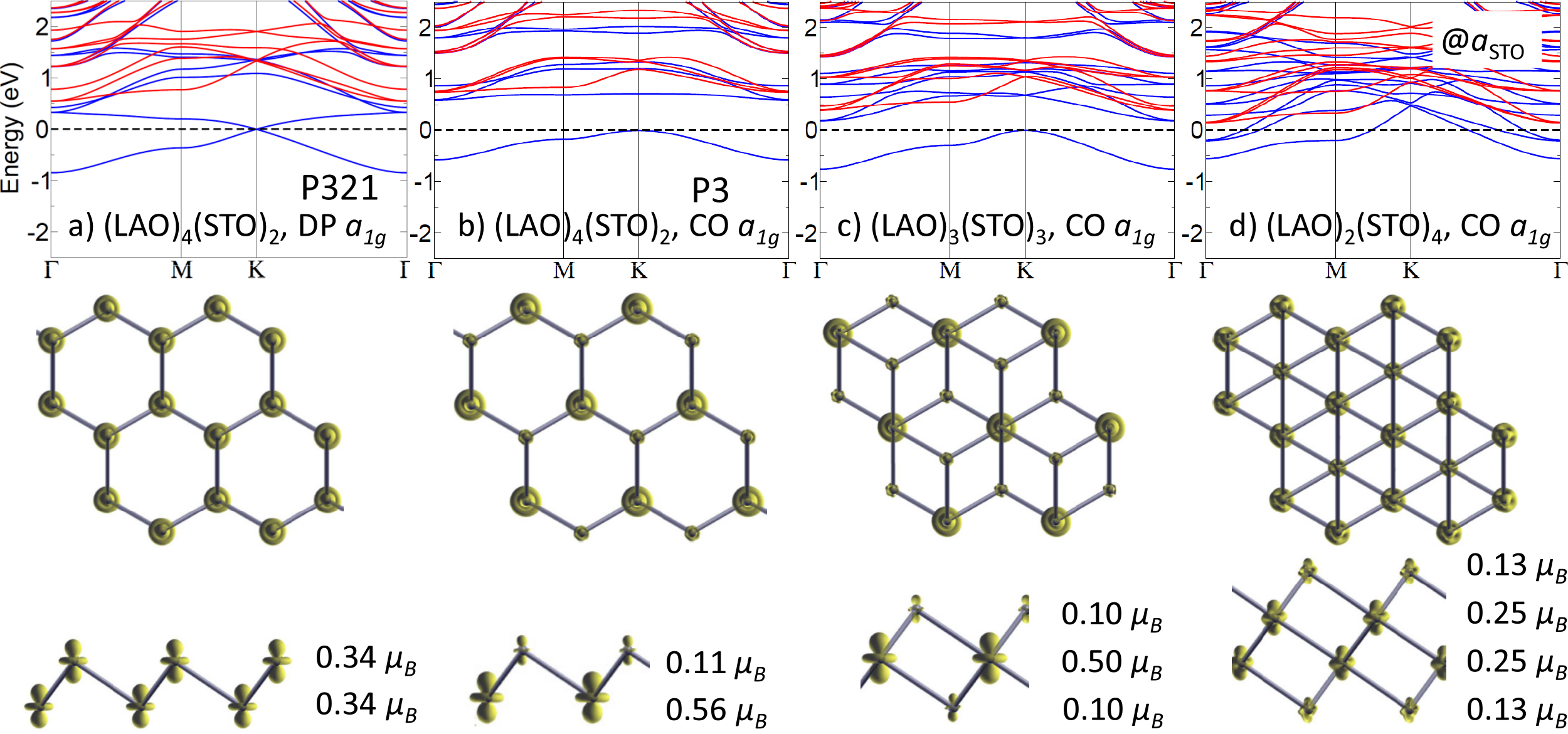}% Here is how to import EPS art
%\vspace{-2.0cm}
\caption{\label{fig:Nsto-asto} As in Fig.~\ref{fig:Nsto-alao}, but for the superlattices at \asto. Note 
the \ag\ occupation independent of $N$. $N=2$: a) For the inversion symmetric interfaces the 
system is Weyl semimetal with Dirac points (DP) at K,K' as in graphene. b) Allowing breaking of inversion symmetry results 
in inequivalent interfaces, \tith versus \tif, leading to the formation of a (111) dipole layer 
and insulating behavior. c-d) with increasing \nst~the system switches from insulating to conducting behavior. Additionally, excess charge is redistributed from the interface Ti layer to the central ones.   }
\end{figure*}

% The goal of this study is to investigate the mechanisms of electronic reconstruction in LAO/STO superlattices with (111) orientation and especially  the possibility to realize exotic, orbitally reconstructed electronic states  and topologically nontrivial phases as a function of strain, crystallographic orientation and thickness of the STO quantum well. 

%As mentioned before we expect the formation of a 2DEG in (111) oriented superlattices due to the charge mismatch at the IF. A further fascinating feature in (111) oriented interfaces is the formation of a buckled honeycomb lattice of the B-site cations. Such a lattice geometry is interesting, because it can support the formation of Dirac points.

%As mentioned before we expect the formation of a 2DEG in (111) oriented superlattices due to the charge mismatch at the IF. A further fascinating feature in (111) oriented interfaces is the formation of a buckled honeycomb lattice of the B-site cations. Such a lattice geometry is interesting, because it can support the formation of Dirac points.

% \textbf{Calculational details.}
DFT calculations have been performed on \lamstn(111) superlattices with varying thickness $M$, $N$ of 
both constituents, using the all-electron full-potential linearized augmented plane wave (FP-LAPW) method, as 
implemented in the WIEN2k code~\cite{wien,wien2k}. The LAO thickness $M$ is always large enough to confine
the carriers to STO. For the exchange-correlation potential we used the generalized 
gradient approximation (GGA) \cite{pbe96}. Static local electronic correlations were taken into account in 
the GGA+$U$ method~\cite{anisimov93} with $U=5$~eV, $J=0.7$~eV (Ti $3d$), $U=8$~eV (La $4f$). As discussed in the Suppl. Material~\cite{suppl}, the obtained solutions are found to be robust with respect to variations of the on-site Coulomb repulsion parameter. The influence of strain was investigated by choosing the lateral lattice parameter of either LAO 
(\alao =3.79~\AA) or STO (\asto =3.92~\AA), which  correspond to superlattices grown either on a LAO(111) or STO(111) substrates. We note that these lateral lattice constants  impose different strain states in the two parts of the superlattice. 
Octahedral tilts and distortions were fully taken into account when relaxing atomic positions, whether constrained
to P321 symmetry (3-fold rotation plus ${\cal I}$)
or fully released to P1 symmetry (in most cases solutions retained the higher P3 symmetry). % since there is no further instability.

% \textbf{Results.}
{\it Results for \alao, corresponding to an underlying LAO(111) substrate.} 
Fig.~\ref{fig:Nsto-alao} presents results for $n$-type (111) oriented 
\lamstn\ superlattices with  thicknesses $N$=2,3,4 of the STO quantum well, each of which is ferromagnetic (FM).   The bilayer at \alao\ (Fig.~\ref{fig:Nsto-alao}a) is a {\it charge ordered FM insulator} with two distinct interfaces with \tith\ (0.60$\mu_B$) and \tif\ (0.10$\mu_B$), respectively,  and due to broken inversion symmetry it is also ferroelectric (FE).  The occupied orbital assumes local \dxy\ orientation (similar to the (001) superlattices~\cite{pentcheva2006,pentcheva2008,zong2008,popovich,XASdata}).
%  and the top of the gap is bounded by a remarkably flat band. 
This state (P1 symmetry) is preferred by 42 meV/Ti over the inversion symmetric case (P321 symmetry, Fig.~\ref{fig:Nsto-alao}b), where in contrast \egp\ orbitals 
become preferentially occupied,  indicating strong competition of
electronic states with distinct orbital occupation, with very different symmetries, and electronic properties (ungapped versus gapped).  \egp\ orbital occupation is preferred also for $N=3$ and 4 (Fig.~\ref{fig:Nsto-alao}c-d)~\cite{cparam}. 
%  ().
The difference $N\rightarrow 2-4$ (Fig.~\ref{fig:Nsto-alao}b-d) in electronic structure seems minor: the flattish lower 
conduction band is mostly occupied,
leaving a hole Fermi 
surface (FS) surrounding the zone corner point K, with charge being balanced by 
one or two electron FS pockets centered at $\Gamma$. For the thicker $N$=3, 4 STO QWs, the extra $e/2$ charge from each interface is 
distributed preferentially towards the central 
layers, related to the different chemical environment of interface vs. central Ti ions.

%\begin{figure}[h!]
%\includegraphics[scale=0.6]{fig4}
%\caption{\label{fig:2sto-asto} For tensile strain (\asto): band structure with top and side view of the density distribution, 
%integrated over occupied Ti $3d$ states, for $2$STO/$4$LAO(111). a) For the inversion symmetric interfaces the 
%system is semimetallic with a Dirac point (DP) at K, as in graphene. Allowing breaking of inversion symmetry results 
%in inequivalent interfaces, \tith versus \tif\, leading to the formation of a (111) dipole layer and insulating behavior.}
%\end{figure}

{\it Results for \asto, corresponding to an underlying STO(111) substrate.} Using the in-plane STO lattice constant strains the LAO layer but
leaves STO cubic (subject to relaxation), it
{\it reverses} the orbital polarization -- a remarkably strong strain effect -- and produces
richer behavior: the symmetric \ag\ orbital becomes occupied 
independently of the \nst\ thickness, as shown in Fig.~\ref{fig:Nsto-asto} (only for $N$=4 the shape is distorted). Similar to the compressive case, 
the charge is shifted from the interface towards the central layers with increasing STO thickness. 
For the FM CO insulating $N=2$ case, the top of the gap is bounded by a remarkably flat band.
For the dice lattice (3/3) case the \tith\ central layer (0.50~\muB) is sandwiched by \tif\ interface
layers, a confinement effect resulting in a FM insulating ground state.
An insulator-to-metal transition occurs at $N=4$, always retaining FM order, although the exchange splitting is reduced with increasing STO QW width. For LAO layers grown on
STO(111), Herranz {\it et al.} found a critical thickness of $\sim$10-12 LAO layers for the onset of
conductivity, but their setup~\cite{Herranz2012} is not comparable to our QW system.
We note from both Figs.~\ref{fig:Nsto-alao} and~\ref{fig:Nsto-asto} the proclivity of linear
``Dirac'' bands to occur at K, but when such points are not
pinned to E$_F$ they have no consequence.

\vskip 1mm
As mentioned, the case $N=2$ is special because the Ti bilayer forms a buckled honeycomb lattice, prompting us to study 
this system in more detail. The single electron can be shared equally and symmetrically by the two Ti ions, or
it can tip the balance to charge order, which requires symmetry breaking from P321 to P3 or possibly P1. 
Both CO and non-CO scenarios can be handled with or without other broken symmetries. We remind that in this \tg\ system we always find the
spin symmetry is broken to FM order, regardless of in-plane strain, restriction of symmetry, and starting configuration. Similarly, 
in their model studies of an $e_g$ bilayer honeycomb lattice (LNO superlattices), 
R\"uegg {\it et al.}\cite{Ruegg2012} found FM ordering
for broad ranges of model parameters.

%{\it STO bilayer at \alao.}
%%%   compressive strain
%%%
% With decreasing \nst the bandwidth at $\Gamma$ decreases slightly,
% which leads to an enhancement of the effective mass, indicating enhancement of electron correlation effects.
%%%   tensile strain
%%%

Constrained to P321 symmetry, a graphene-like Dirac point emerges at the zone corner point K that is pinned
to the Fermi level and protected
by the equivalence of the Ti sites, see 
Fig.~\ref{fig:Nsto-asto}a. True particle-hole symmetry is restricted to relatively low energy due to coupling of the
upper band to high-lying bands.  
The occupied bandwidth corresponds to hopping $t_{a_{1g},a_{1g}}$=0.28 eV.
Having a single electron shared symmetrically by two Ti sites is potentially unstable.
Allowing breaking of this ${\cal I}$ symmetry results again, as for compressive strain, 
in a CO \tith\ (0.56$\mu_B$) and \tif\ (0.11$\mu_B$), FM, FE, and insulating state evident from
Fig.~\ref{fig:Nsto-asto}b that is 0.95~meV/Ti more stable. The massless-to-massive transformation of the spectrum results from breaking of the equivalence
of the Ti ions, {\it i.e.} ${\cal I}$ symmetry. Nearly complete charge disproportionation gives CO alternating around each honeycomb hexagon,
retaining the P3 symmetry that stabilizes the system. 

In both cases (Fig. \ref{fig:Nsto-alao}a and ~\ref{fig:Nsto-asto}b) charge ordering
is accompanied by formation of an electric dipole in the bilayer as well as Ti-O bond alternation: 
in the latter tensile case the Ti-Ti interlayer distance is 2.1\AA, with  
\tif-O (\tith-O) distance of 1.91\AA\ (1.98\AA), respectively (being 1.94\AA~in the symmetric case). 
The orbital polarization, which is pure \ag\ occupation, is distinct from the CO state for compressive strain,
which has \dxy\ character. 
The complex occupied orbitals that arise in QWs of SrVO$_3$\cite{SVOSTO_Pardo} and VO$_2$\cite{VO2}
are not evident in the present system.

The occupied bandwidths  and shapes of the CO states for the two strain states are very similar, and in each case the gap of
$\sim$0.8~eV leads to a higher-lying remarkably flat band (Fig.~\ref{fig:Nsto-asto}b), 
which for tensile strain forms the top of the gap into which doped
electrons would go. 
Nearly flat bands are interesting in the context of fractional quantum Hall effect. Flat bands have been obtained in a variety of cases: from a  
$d$-$p_x$-$p_y$ square lattice model~\cite{Sun2011} and from a $p_x$,$p_y$ honeycomb lattice
model~\cite{Wu2007};
Xiao {\it et al.}\cite{Xiao2011}  
and Fiete and collaborators\cite{Ruegg2012,Hu2012} found that perfectly flat bands emerge also in an 
\eg\ bilayer model, and R\"uegg {\it et al.} demonstrated~\cite{Ruegg2012} that they arise 
from strictly localized eigenstates with symmetric orbital ordering; several other examples have been discovered
and studied.\cite{Bergman2008,Weeks2012,Wang2011,Hou2009,Green2010,Katsura2010}  

The band structure we obtain is qualitatively different to the case of LNO ($e_g$) superlattices where the bands close to the Fermi level consist of  four bands: two linearly  crossing bands and two flat bands with a quadratic touching point at $\Gamma$~\cite{Yang2011,Ruegg2012}. In contrast, for this \tg\ system at low filling there are only two relevant bands, and inversion symmetry breaking gaps the two linearly crossing bands into two relatively flat bands.  The very flat conduction band we find
occurs only for the CO states (which are the ground states) that involve either \ag\ and \dxy\ orbital occupation, depending on strain. 
Its origin therefore is not as straightforward as for the above mentioned models.

Allowing a difference in on-site $3d$ energies will be an essential part of breaking ${\cal I}$ symmetry when 
modeling the transition from massless Dirac pair
into massive (gapped) bands, see Fig.~\ref{fig:Nsto-asto}a,b. This difference in \tif\ and \tith\ $3d$ 
energies will be similar to the $2p$
core level difference arising from different Ti-oxygen bond lengths, which at 1.7 eV is {\it twice} the occupied bandwidth of the
Dirac band structure.  Focusing on the majority (blue) bands of the CO state, a regularity can be
seen: there is a parallel pair of bands of the same shape as the occupied band, but lying 1.3 eV 
higher, which is the pair of \tith\ \egp\ bands. Mirroring the flat band and again 1.3 eV
higher are two more (no longer precisely) flat bands; these are the \tif\ \ag\ and \egp\ bands.
% A result of the charge ordering is that the occupied  bandwidth decreases
% from 0.9 eV to 0.6 eV.

As our analysis shows~\cite{suppl}, both the Dirac point and the charge ordered state result from on-site correlation. Both solutions emerge at relatively small $U$ values (1-2 eV) and are robust with respect to further increase of $U$. In the CO case, the primary effect of $U$ is to encourage integer orbital occupations, {\it i.e.} a Mott insulating state: four quarter-filled spin-orbitals (two sites, two orbitals) convert by CO to one empty sublattice
and one half-filled sublattice, which then becomes Mott insulating.
This CO-Mott transition is driven
by a combination of Hubbard $U$ and symmetry breaking (accompanied by a substantial
oxygen relaxation), and may be aided by intersite repulsion. 

Bands on the same graphene lattice, displaying the topology that we find for the Dirac bilayer, have been related
 to topological character.\cite{Xiao2011,Ruegg2011,Yang2011} 
%Our first principles methods are not well suited to calculate the edge states to identify topological
%character (the required supercell would require 10-20 times more atoms at least), but topological
%character is plausible.  
Typically topological phases are protected by time reversal symmetry coupled with additional symmetries. 
Fu~\cite{FuCTI} and others~\cite{FieteCTI,Slager2013} have noted that SOC is not required, demonstrating
that topological band insulators can also be protected by crystalline symmetry rather than ${\cal T}$ symmetry. 
%{bf P1 U parameter drives the system between a metal and a trivial Mott insulator, but for P321
%with U Dirac point driven to Ef but further increase  
%parity at TRIM points in BZ (Hasan review) odd  -1, even +1 have to be mulitplied; if -1 TI
%magnetic top Insul Savrasov axion PRL 2012

%In all of the present cases FM order
%breaks time reversal symmetry, so topological character becomes progressively less likely.

%\begin{figure*}[h!]
%\includegraphics[scale=0.5]{2STO4LAO@LAO_uu_ud.pdf}% Here is how to import EPS art
%\caption{\label{fig:Nsto-asto-2}Band structure and occupied density distribution,   $2$STO/$4$LAO(111) under 
%compression (\alao), showing the charge ordered FM insulator.  Note the ``$d_{xy}$'' occupation, {\it i.e.}
%single (but unsymmetrically oriented) real $t_{2g}$ orbital.}
%\end{figure*}

%The various phases we have discussed have broken all possible symmetries except 3-fold rotation with a single \tith-\tif\ distance.
%Specifically, no \tith-\tif\, nor Ti$^{3.5+}$-Ti$^{3.5+}$, dimerization
%has been obtained. The latter could promote a triangular lattice version of the 
%dimer Mott insulator (DMI),\cite{Fukuyama0,Ogata2006} a phase that is
%is known to compete with charge ordering and loses in this bilayer. The DMI ground state appears to account for the
%metal-insulator transition in quasi-2D organic crystals\cite{Fukuyama0} and has been suggested to
%arise in a digital oxide nanostructures,\cite{ChenLeeBalents} and a trimer Mott insulating phase 
%accounts for MIT in the insulating 2D nickelate La$_4$Ni$_3$O$_8$ with average Ni$^{4/3+}$ charge 
%state.\cite{La428}

Now we summarize. Unexpected richness has been uncovered in (111)-oriented STO/LAO heterostructures,
where carriers must reside in Ti \tg\ states. The competing ground states are ferromagnetic, with strain-controlled crystal
field splitting $t_{2g} \rightarrow$ \ag\ + \egp\, promoting strain engineering of orbital polarization.
For the system under tensile strain, a graphene-like Dirac point degeneracy survives as long
as inversion symmetry of the bilayer is preserved. Allowing breaking of this symmetry, charge ordering
with a flat conduction band and multiferroic properties results, again with orbital polarization
dependent on strain.  Melting of the CO phase as temperature is raised, where several symmetries
(and conductivity) are restored, should reveal very rich behavior.

\begin{acknowledgments}
R.P. acknowledges discussions with M. Rozenberg. R.P. and D. D. acknowledge financial support through the DFG SFB/TR80  and  grant {\sl h0721} for computational time at the Leibniz Rechenzentrum.  W. E. P. was supported by U.S. Department of Energy Grant No. DE-FG02-04ER46111.
\end{acknowledgments}


\begin{thebibliography}{99}

\bibitem{Hwang2012}
%  Emergent phenomena at oxide interfaces
  H. Y. Hwang, Y. Iwasa, M. Kawasaki, B. Keimer, N. Nagaosa, and Y. Tokura,
  Nat. Mater. {\bf 11}, 103 (2012).

\bibitem{Huijben2009} M. Huijben, A. Brinkman,
  G. Koster, G. Rijnders, H. Hilgenkamp, and D. A. Blank Adv. Mater. {\bf 21}, 1665 (2009). 
  
 \bibitem{Mannhart} J. Mannhart and D. G. Schlom,
 Science {\bf 327}, 1607 (2010).
 
\bibitem{jpcm2010} R. Pentcheva and W. E. Pickett,
 J. Phys.: Condens. Matter {\bf 32}, 043001 (2010).

\bibitem{zubko} P. Zubko, S. Gariglio,
  M. Gabay, P. Ghosez, and J.-M. Triscone,  Annu. Rev. Condens. Matter Phys. {\bf 2}, 141 (2011).
 
 \bibitem{overview} R. Pentcheva, R. Arras, K. Otte, V. G. Ruiz, and W. E. Pickett,
 Phil. Trans. R. Soc. A {\bf 370}, 4904 (2012).

\bibitem{Herranz2012}
% arXiv:1210.7955
%  High mobility conduction at (110) and (111) LaAlO3/SrTiO3 interfaces
  G. Herranz, F. Sanchez, N. Dix, M. Scigaj, and J. Fontcuberta,
   Scientific Reports (Nature group) {\bf 2}, 758 (2012).

\bibitem{Haldane88}F. D. M. Haldane, Phys. Rev. Lett. {\bf 61}, 2015 (1988).
% Model for a Quantum Hall Effect without Landau Levels: Condensed-Matter Realization of the "Parity Anomaly"

% #5
\bibitem{Xiao2011}
% arXiv:1106.4296 [pdf, ps, other]
% Interface engineering of quantum Hall effects in digital transition metal oxide heterostructures
   D. Xiao, W. Zhu, Y. Ran, N. Nagaosa, and S. Okamoto,
   Nature Commun. {\bf 2}, 596 (2011).

\bibitem{Ruegg2011}
% Electronic structure of (LaNiO3)2/(LaAlO3)N heterostructures grown along [111]
A. R\"uegg and G. A. Fiete,
Phys. Rev. B {\bf 84}, 201103 (2011).


\bibitem{Yang2011}
% arXiv:1109.1551
%  Possible interaction driven topological phases in (111) bilayers of LaNiO3
  K.-Y. Yang, W. Zhu, D. Xiao, S. Okamoto, Z. Wang, and Y. Ran,
  Phys. Rev. B {\bf 84}, 201104(R) (2011).


\bibitem{Ruegg2012}
% Electronic structure of (LaNiO3)2/(LaAlO3)N heterostructures grown along [111]
A. R\"uegg, C. Mitra, A. A. Demkov, and G. A. Fiete,
Phys. Rev. B {\bf 85}, 245131 (2012).

\bibitem{Gilbert2012} M. Gilbert, P. Zubko, R. Scherwitzl, and J.-M. Triscone, Nat.
   Mater. {\bf 11}, 195 (2012).
%  LNOLMO111 growth

\bibitem{Middey2012}
% arXiv:1212.0590
%  Epitaxial growth of (111)-oriented LaAlO$_3$/LaNiO$_3$ ultra-thin superlattices
  S. Middey {\it et al.}, 
%    D. Meyers, M. Kareev, E. J. Moon, B. A. Gray, X. Liu, J. W. Freeland, and J. Chakhalian,
  Appl. Phys. Lett. {\bf 101}, 261602 (2012).
%  arXiv:1212.0590. APL published Dec 2012

\bibitem{pentcheva2006}
   R. Pentcheva and W. E. Pickett,
%    "Charge localization or itineracy at LaAlO$_3$/SrTiO$_3$ interfaces:
%     Hole polarons, oxygen vacancies, and mobile electrons",
   Phys. Rev. B {\bf 74}, 035112 (2006).

\bibitem{pentcheva2008}
   R. Pentcheva and W. E. Pickett,
% "Ionic relaxation contribution to the electronic reconstruction at the $n$-type LaAlO$_3$/SrTiO$_3$ interface",
   Phys. Rev. B {\bf 78}, 205106 (2008).

\bibitem{zong2008}Z. Zhong and P. J. Kelly, Eur. Phys. Lett. {\bf 84}, 27001, (2008).
\bibitem{popovich}Z. S. Popovic, S. Satpathy, and R. M. Martin, 
  Phys. Rev. Lett. {\bf 101}, 256801 (2008).
%%Phys. Rev. Lett. {\bf 94}, 176805 (2005).

\bibitem{XASdata} M. Salluzzo {\it et al.},
  Phys. Rev. Lett. {\bf 102}, 166804 (2009).
%  M. Salluzzo, J. C. Cezar, N. B. Brookes, V. Bisogni, G. M. De Luca, C. Richter,
%  S. Thiel, J. Mannhart, M. Huijben, A. Brinkman, G. Rijnders, and G. Ghiringhelli,


\bibitem{SVOSTO_Pardo}V. Pardo and W. E. Pickett, Phys. Rev. B {\bf 81}, 245117 (2010).
% Electron Confinement, Orbital Ordering, and Orbital Moments in d0-d1 Oxide Heterostructures   

\bibitem{wien}
   K. Schwarz and P. Blaha, Comp. Mat. Sci. {\bf 28}, 259 (2003).
\bibitem{wien2k}
    P. Blaha, K. Schwarz, G. K. H. Madsen, D. Kvasnicka, and J. Luitz,
{\it WIEN2k, An Augmented Plane Wave Plus Local Orbitals Program for Calculating Crystal Properties}, 
   ISBN 3-9501031-1-2
    (Vienna University of Technology, Vienna, Austria, 2001).

\bibitem{pbe96}J. P. Perdew, K. Burke, and M. Ernzerhof,
  Phys. Rev. Lett. {\bf 77}, 3865 (1996).
\bibitem{anisimov93}V. I. Anisimov, J. Zaanen, and O. K. Andersen, (1993).
\bibitem{suppl} See Supplemental Material at http://link.aps.org/supplemental/... for details on lattice relaxations and dependence on $U$.
\bibitem{cparam}The orbital polarization is quenched to \tg\ only for strong compression of the $c$ lattice constant%  \bibitem{pentcheva2007}
%     R. Pentcheva and W. E. Pickett,
%%% "Correlation-driven charge order at the interface between a Mott insulator and a band insulator",
%     Phys. Rev. Lett. {\bf 99}, 016802 (2007).

% \bibitem{siemons}
%     W. Siemons {\it et al.}, 
%%% G. Koster, H. Yamamoto, W. A. Harrison, G. Lucovsky, T. H. Geballe, 
%        D. H. A. Blank, and M. R. Beasley,
%%%  "Origin of charge density at LaAlO$_3$ on SrTiO$_3$ Heterointerfaces: Possibility of intrinsic doping",
%     Phys. Rev. Lett. {\bf 98}, 196802 (2007).

\bibitem{VO2} V. Pardo and W. E. Pickett, Phys. Rev. B {\bf 81}, 035111 (2010).
%   Metal-insulator Transition Through a semi-Dirac Point in Oxide Nanostructures:
%   VO2 (001) Layers Confined Within TiO2

\bibitem{Sun2011}
% Nearly Flatbands with Nontrivial Topology
K. Sun, Z. Gu, H. Katsura, and S. Das Sarma,
Phys. Rev. Lett. {\bf 106}, 236803 (2011).

\bibitem{Wu2007}
% Flat Bands andWigner Crystallization in the Honeycomb Optical Lattice
  C. Wu, D. Bergman, L. Balents, and S. Das Sarma, Phys. Rev. Lett. {\bf 99}, 070401 (2007).

\bibitem{Hu2012}
% arXiv:1211.0562 [pdf, ps, other]
%   Topological phases in layered pyrochlore oxide thin films along the [111] direction
  X. Hu, A. R\"uegg, and G. A. Fiete,
  Phys. Rev. B {\bf 86}, 235141 (2012).

\bibitem{Bergman2008}
% Band touching from real-space topology in frustrated hopping models
D. L. Bergman, C. Wu, and L. Balents,
Phys. Rev. B {\ 78}, 125104 (2008).

\bibitem{Weeks2012}
% Flat bands with nontrivial topology in three dimensions
C. Weeks and M. Franz,
Phys. Rev. B {\bf 85}, 041104 (2012).


\bibitem{Wang2011}
% Nearly flat band with Chern number C=2 on the dice lattice
F. Wang and Y. Ran,
Phys. Rev. B {\bf 84}, 241103(R) (2011).

\bibitem{Hou2009}
% Unconventional Energy Bands and Chirality of Ultracold Atoms in Trilayer Honeycomb Lattice
J.-M. Hou,
Commun. Theor. Phys. {\bf 52}, 247 (2009).

\bibitem{Green2010}
% Isolated flat bands and spin-1 conical bands in two-dimensional lattices
D. Green, L. Santos, and C. Chamon,
Phys. Rev. B {\bf 82}, 075104 (2010).


\bibitem{Katsura2010}
% Ferromagnetism in the Hubbard model with topological/non-topological flat bands
H. Katsura, I. Maruyama, A. Tanaka, and H. Tasaki,
EPL (Europhysics Letters) {\bf 91}, 57007 (2010).

\bibitem{FuCTI} L. Fu, Phys. Rev. Lett. {\bf 106}, 106802 (2011).
% Topological crystalline insulators.

\bibitem{FieteCTI}M. Kargarian and G. A. Fiete, Phys. Rev. Lett. {\bf 110}, 156403 (2013).
% Topological Crystalline Insulators in Transition Metal Oxides


\bibitem{Slager2013}
% arXiv:1209.2610
%  The space group classification of topological band insulators
   R.-J. Slager,   A. Mesaros, V. Juricic, and J. Zaanen,
    Nat. Phys. {\bf 9}, 98 (2013).

%\bibitem{Fukuyama0}H. Kino and H. Fukuyama, J. Phys. Soc. Japan 
%  {\bf 64}, 1877 (1995).
%\bibitem{Ogata2006}H. Seo, J. Merino, H. Yoshioka, and M. Ogata,
%  J. Phys. Soc. Japan {\bf 75}, 051009 (2006).
%\bibitem{ChenLeeBalents}R. Chen, S.-B. Lee, and L. Balents, arXiv:1301.4222.
%  dimer Mott insulator in oxide nanostructures
%\bibitem{La428}V. Pardo and W. E. Pickett, Phys. Rev. Lett. {\bf 105}, 266402 (2010).
%% Quantum Confinement Induced Molecular Mott Insulating State in La4Ni3O8 

% arXiv:1302.3253 [pdf, ps, other]
%   Quantum Confinement Induced Magnetism in LaNiO$_3$-LaMnO$_3$ Superlattices
%   Shuai Dong, Elbio Dagotto 

\end{thebibliography}
\end{document}